\documentclass[twocolumn,english,showpacs,preprintnumbers,amsmath,amssymb,floatfix]{revtex4}
\usepackage[T1]{fontenc}
\usepackage{color}
\usepackage{array}
\usepackage{amstext}
\usepackage{graphicx}
\usepackage{esint}

\makeatletter

\@ifundefined{textcolor}{}
{%
 \definecolor{BLACK}{gray}{0}
 \definecolor{WHITE}{gray}{1}
 \definecolor{RED}{rgb}{1,0,0}
 \definecolor{GREEN}{rgb}{0,1,0}
 \definecolor{BLUE}{rgb}{0,0,1}
 \definecolor{CYAN}{cmyk}{1,0,0,0}
 \definecolor{MAGENTA}{cmyk}{0,1,0,0}
 \definecolor{YELLOW}{cmyk}{0,0,1,0}
 }

\@ifundefined{definecolor}
 {\usepackage{color}}{}
\@ifundefined{definecolor}
 {\usepackage{color}}{}
\makeatother

\makeatother

\usepackage{babel}
\begin{document}

\title{Spin-dependent energy distribution of B-hadrons from
polarized top decays considering the azimuthal correlation rate}

\author{S. M. Moosavi Nejad$^{a,b}$}

\email{mmoosavi@yazduni.ac.ir}

\affiliation{$^{(a)}$Faculty of Physics, Yazd University, P.O. Box
89195-741, Yazd, Iran}

\affiliation{$^{(b)}$School of Particles and Accelerators,
Institute for Research in Fundamental Sciences (IPM), P.O.Box
19395-5531, Tehran, Iran}

\date{\today}

\begin{abstract}

In our previous work, we studied the  polar  distribution of the scaled energy 
of bottom-flavored hadrons from polarized top quark decays $t(\uparrow)\rightarrow W^++b(\rightarrow X_b)$,
using two different helicity coordinate systems. 
Basically, the energy distributions are governed by the  unpolarized, polar and azimuthal rate functions 
which are related to the density matrix elements of the decay $t(\uparrow)\rightarrow W^++b$.
Here we present, for the first time, the analytical expressions for the 
${\cal O}(\alpha_s)$ radiative corrections to the differential azimuthal  decay rates of the 
partonic process $t(\uparrow)\rightarrow b+W^+(+g)$ in two helicity systems, which are needed to study the azimuthal
distribution of the energy spectrum of the B-hadron produced in polarized top quark decays.
Our predictions of the hadron energy distributions  enable us to deepen our knowledge 
of the hadronization process and to determine the polarization states of top quarks.

\end{abstract}

\maketitle

\section{Introduction}

In the Standard Model (SM), the top quark has a short lifetime ($\approx 0.5\times 10^{-24} s$ \cite{Chetyrkin:1999ju})
so decays rapidly and this short time does not allow the top quark to form the QCD bound states, phrased in a
different language, its short lifetime  implies that it decays before hadronization takes place.
If it was not for the confinement of color, the top quark could be considered as a free particle and 
this property allows it to behave like a real particle and one can safely describe its decay in perturbative theory.
In fact, at the top mass scale the strong coupling constant is small, $\alpha_s(m_t)\approx 0.107$, so that all QCD effects involving
the top quark are well behaved in the perturbative sense.
Due to the Cabibbo-Kobayashi-Maskawa (CKM) mixing matrix 
element $V_{tb}=0.999$ \cite{Cabibbo:1963yz}, the decay width of the top quark is 
almost exclusively dominated by the two-body 
channel at the lowest order where a W-gauge boson and a bottom quark are contributed.
As it is well known, bottom quarks produced hadronize before they decay ($b\rightarrow X_b$), 
therefore each $X_b$-jet  contains a bottom-flavored hadron
 which most of the times is a B-meson. The bottom hadronization is indeed one of the largest 
sources of uncertainty in the measurement of the mass of top quark at 
the CERN Large Hadron Collider (LHC) \cite{M.Beneke} and the Tevatron \cite{Abulencia:2005ak}, 
as it contributes to the Monte Carlo systematics.
The LHC is  a superlative top factory, which  allows us to carry out precision tests of the SM
and, specifically, a precise measurement of the top quark properties such as its  mass $m_t$, total decay width $\Gamma_t$
and branching fractions.
At the LHC, of particular interest is the distribution in the energy of meson produced in the top quark rest frame,
so that this energy distribution provides direct access to the bottom fragmentation functions (FFs).
In \cite{Kniehl:2012mn}, we studied both the  B-meson energy
distribution produced from unpolarized top decay and we studied  the angular distribution
of the W-boson decay products in the decay chain $t\rightarrow bW^+\rightarrow Bl^+\nu_l+X$. \\
Since the top quark decays rapidly so that its life time scale is much shorter than the typical time needed for the QCD 
interactions to randomize its spin, therefore its full polarization content is 
preserved when it decays and passes on to its decay products. 
Hence, the polarization of the top quark will reveal itself in the angular 
decay distribution and can be studied through the angular correlations 
between the direction of the top quark spin and the momenta of the decay products, $W^+$-boson and $b-$quark. 
In \cite{Nejad:2013fba}, we studied the ${\cal O}(\alpha_s)$ angular distribution of the  scaled 
energy of the B-hadrons, by calculating the polar angular correlation in the rest frame decay of a polarized top quark 
into a stable $W^+$-boson and B/D-hadrons. We analysed this angular correlation in a helicity coordinate system (system~1) where
the event plane, including the top and its decay products, is defined in the $(x, z)$ plane with 
the z-axes along the bottom quark momentum.
In this system the top polarization vector was evaluated with respect to the direction of the bottom quark momentum. 
Basically, to define the planes we need to measure the momentum directions of the $\vec{p}_b$ and $\vec{p}_W$ 
and the polarization direction of the top quark, 
where the evaluation of the momentum direction of $\vec{p}_b$ requires the use of a 
jet finding algorithm, whereas the top spin direction must be obtained from the theoretical input. For example,
in $e^+e^-$ interactions the polarization degree of the top can be tuned with the help of polarized beams \cite{Parke},
so that a polarized linear $e^+e^-$ collider may be considered as a copious source of close to zero and close to $100\%$ polarized tops.\\
In \cite{Nejad:2014sla}, we analysed the ${\cal O}(\alpha_s)$ polar distribution of the B-hadron energy 
in a different helicity coordinate system (system~2) where, as in \cite{Nejad:2013fba} the event plane is the $(x, z)$ plane 
but with the z-axes along the $W^+$-boson , so the polarization direction
of the top quark is evaluated with respect to the $W^+$ momentum vector. This election makes the
calculation so complicated. \\
The azimuthal correlations between the  event plane and the intersecting ones to this plane evaluated
in two helicity systems belong to a class of polarization observables involving
the top quark in which the leading-order (LO) contribution gives a zero result in the SM, so
the non-zero contributions can either arise from higher order SM radiative corrections
or non-SM effects \cite{Groote:2006kq}.
Since highly polarized top quarks will become available at hadron colliders
through single top production processes, which occur at the $33\%$ level of the $t\bar{t}$ pair production rate \cite{Mahlon:1996pn},
it will then be possible to experimentally measure the azimuthal correlation 
between the $(\vec{p}_W, \vec{p}_b)$ and $(\vec{P}_t, \vec{p}_b)$ planes
in the helicity system~1 and the $(\vec{p}_W, \vec{p}_b)$ and $(\vec{P}_t, \vec{p}_W)$ planes
in the helicity system~2. Here, $\vec{P}_t$ stands for the polarization vector of the top quark and
$\vec{p}_W$ and $\vec{p}_b$ stand for the four-momenta of W boson and bottom quark, respectively.
To analyse the aforementioned azimuthal correlations in the polarized top rest frame, we study
the azimuthal distribution of the scaled energy of B-hadrons at the process 
$t(\uparrow)\rightarrow W^++B+X$ at NLO, by calculating 
the azimuthal decay distribution of a polarized top quark in the partonic process
$t(\uparrow)\rightarrow b+W^+(+g)$ in two aforementioned coordinate systems.
For the nonperturbative part of the process ($b, g\rightarrow B+X$),
from Ref.~\cite{Kniehl:2008zza} 
we apply the realistic  $(b, g)\rightarrow B$ FFs
obtained through a global fit to
$e^+e^-$ data from CERN LEP1 and SLAC SLC.\\
Finally, we shall present and compare our numerical results in both systems.
These measurements will be important to deepen our understanding of the 
nonperturbative aspects of B-hadrons  formation  and to test the universality and scaling violations of the B-hadron FFs.

This paper is structured as follows.
In Sec.~\ref{sec:one}, we introduce the angular rate structure by defining 
the technical details of our calculations.
In Sec.~\ref{sec:two}, our analytic results for the ${\cal O}(\alpha_s)$ QCD corrections to the 
azimuthal distributions of partial decay rates are presented.
In Sec.~\ref{sec:three}, we shall make our predictions of energy distribution
of B-hadrons and present our numerical analysis.
In Sec.~\ref{sec:four},  our conclusions are summarized.

\section{Angular structure of partial decay rate}
\label{sec:one}

In the current-induced $t\rightarrow b$ transition, the dynamics of the process is embodied in the hadron tensor
$H^{\mu\nu}\propto \sum_{X}\left\langle t|J^{\mu\dagger}|X\right\rangle\left\langle X|J^\nu|t\right\rangle$, where 
the SM  current combination  is given by $J_\mu=J_\mu^V-J_\mu^A$.
Here, the left-chiral components of the weak current are  given by 
$J_\mu^V\propto\bar\psi_b \gamma_\mu\psi_t$ and $J_\mu^A\propto\bar\psi_b \gamma_\mu\gamma_5\psi_t$.
In the transition $t\rightarrow W^++b(+g)$, the intermediate states are $|X>=|b(p_b)>$ for the 
Born term and virtual contributions and $|X>=|b+g>$ for the ${\cal O}(\alpha_s)$ real contributions.

The general angular distribution of the differential decay width $d\Gamma/dx$ of a polarized top quark 
decaying into a jet $X_b$ with bottom quantum  numbers and a $W^+$ boson is expressed  by the following
 form 
\begin{eqnarray}\label{form}
\frac{d\Gamma}{dx_id\cos\theta_P d\phi_P}&=&\frac{1}{4\pi}\bigg\{\frac{d\Gamma_A}{dx_i}+P\frac{d\Gamma_B}{dx_i}\cos\theta_P\nonumber\\
&&+P\frac{d\Gamma_C}{dx_i}\sin\theta_P\cos\phi_P\bigg\},
\end{eqnarray}
where the polar and azimuthal angles $\theta_P$ and $\phi_P$ show the orientation of
the plane including the spin of the top quark relative to the event plane (see \cite{Nejad:2014sla}) and $P$ is the
magnitude of the top quark polarization, so $P=0$ stands for an unpolarized top quark while $P=1$
corresponds to $100\%$ top quark polarization. In the notation above, $d\Gamma_A/dx$ corresponds to the unpolarized 
differential decay rate, while $d\Gamma_B/dx$ and $d\Gamma_C/dx$ describe the polar and 
azimuthal correlation between the polarization of the top quark and its decay products, respectively.\\
We shall closely follow the notation of \cite{Kniehl:2012mn}, where the partonic scaled energy fraction $x_i$
is defined as 
\begin{eqnarray}
x_i=\frac{2p_i\cdot p_t}{m_t^2}.
\end{eqnarray}
As we demonstrated in \cite{Kniehl:2012mn}, the finite-$m_b$ corrections are rather small and thus 
to study the scaled energy distributions of the B-meson, we employ  
the massless scheme or  zero-mass variable-flavor-number (ZM-VFN) scheme \cite{jm} in the top quark rest frame, where 
the zero mass parton approximation is also applied  to the bottom quark. The non-zero value of the b-quark mass only enter
through the initial condition of the nonperturbative FFs. Nonperturbative FFs are
describing the hadronization processes $(b, g)\rightarrow X_b$ and are subject to 
Dokshitzer-Gribov-Lipatov-Alteralli-Parisi (DGLAP)  evolution \cite{dglap}.\\
By the zero mass approximation, one has $0\leq x_i\leq 1-\omega$ where $\omega=m_W^2/m_t^2$.
Throughout this manuscript, we apply the normalized partonic energy fraction as
\begin{eqnarray}\label{variable}
x_i=\frac{2E_i}{m_t(1-\omega)}, \qquad (i=b, g)
\end{eqnarray}
where $E_i$ refers to the energy of outgoing partons (bottom or gluon) and $0\leq x_i\leq 1$.

In our previous works, the NLO radiative corrections to the unpolarized 
differential rate $d\Gamma_A/dx_i$ \cite{Kniehl:2012mn} and the polar 
differential rates $d\Gamma_B/dx_i$ \cite{Nejad:2013fba,Nejad:2014sla} have been studied in two possible helicity systems, extensively.
In the present work, we study the radiative corrections to the azimuthal correlation function 
$d\Gamma_C/dx_i$ in both helicity systems, which have not been done before.
Finally, at the hadron level we shall compare our predictions for the energy distribution of B-mesons
in two coordinate systems~1 and 2, considering all contributions.

\section{Analytic results for azimuthal decay distributions}
\label{sec:two}

In the rest frame of a top quark decaying into a $W^+$ boson, a b-quark and a gluon, the final state particles
define an event plane. Relative to this plane one can then define the spin direction
of the polarized top quark. There are two different choices of possible coordinate systems relative to the event
plane where one differentiates between helicity systems according to the orientation of the $z$-axis (these
systems compared in \cite{Nejad:2014sla}). Four-momenta of the b-quark and the $W^+$ boson in these two various coordinate 
systems are defined as
\begin{eqnarray}
System~1&:& \vec{p}_b || z \quad; (\vec{p}_W)_x\geq 0 \nonumber\\
System~2&:& \vec{p}_W || z \quad; (\vec{p}_b)_x\geq 0
\end{eqnarray} 
Indeed, in the system~1 the three-momentum of the b-quark
points into the direction of the positive $z$-axis and in the system~2, the three-momentum of the $W^+$ boson is defined
along this axis.\\
In the following, we explain the technical details of our calculation for the NLO
radiative corrections to the tree-level decay rate of $t(\uparrow)\rightarrow b+W$.

\subsection{Born term results}
\label{sec:two-1}

In the SM, the polarized top decay rate is dominated by the decay 
process $t(p_t)\rightarrow b(p_b)+W^+(p_W)$ at the Born level.
In the rest frame of the top quark, the four-momentum of the top quark
is set to $p_t=(m_t; \vec{0})$ and the polarization four-vector of the top quark
is set as $s_t=P(0; \sin\theta_P\cos\phi_P, \sin\theta_P\sin\phi_P, \cos\theta_P)$, where
$P$ is the top polarization degree ($0\leq P\leq 1$).
Considering the coordinate system~1, where the three-momentum of the b-quark
points into the positive $z$-axis, 
we set the four-momentum of the b-quark  as $p_b=E_b(1; 0, 0, 1)$ and 
in the system~2, it is $p_b=E_b(1; 0, 0, -1)$ where the three-momentum of the $W^+$ boson is defined
along the positive $z$-axis. Note that by applying
the ZM-VFN scheme we put the b-quark mass to zero throughout this paper. Thus, 
the Born term helicity structure of partial rates, reads
\begin{eqnarray}\label{aziz}
\frac{d^2\Gamma^{\textbf{(0)}}}{d\cos\theta_{P}d\phi_P}&=&\frac{1}{4\pi}\bigg\{\Gamma_A^{\textbf{(0)}}\mp 
P\Gamma_B^{\textbf{(0)}}\cos\theta_{P}\nonumber\\
&&+P\Gamma_C^{\textbf{(0)}}\sin\theta_P\cos\phi_P\bigg\},
\end{eqnarray}
where  the sign $''-''$ stands for the helicity system~1 and the sign $''+''$ is for the second one, and 
\begin{eqnarray}
\Gamma_A^{\textbf{(0)}}&=&\frac{m_t \alpha}{16\omega \sin^2\theta_W}(1+2\omega)(1-\omega)^2,\nonumber\\
\Gamma_B^{\textbf{(0)}}&=&\frac{m_t \alpha}{16\omega \sin^2\theta_W}(1-2\omega)(1-\omega)^2, \nonumber\\
\Gamma_C^{\textbf{(0)}}&=& 0و
\end{eqnarray}
where, $\theta_W$ is the weak mixing angle and $\alpha$ is the tiny structure constant.
These results are in complete agreement with the expressions in \cite{Fischer:2001gp,Fischer:1998gsa}.\\
As it is seen, the Born term contribution to $\Gamma_C$ is zero. We 
point out that the vanishing of this azimuthal correlation is a consequence of the
left-chiral (V-A)(V-A) nature of the current-current interaction in the SM. Another example of a LO
zero polarization observable is given in \cite{Kniehl:2012mn}. There, we showed that
the decay of a top quark into a polarized transverse-plus W boson and a (massless)
bottom quark leads to the contribution zero for the top decay rate into the transverse-plus W boson at the Born 
term level due to the left-chiral (V-A) coupling structure of the SM. However, if one takes a massive b-quark 
in the calculation, this contribution
is no longer zero but the LO result obtained for $\Gamma_C$ does not depend on the mass of the
bottom quark.

\subsection{QCD NLO contribution to the azimuthal differential decay rate $d\Gamma_c/dx_i$}
\label{sec:two-2}

Generally, the required ingredients for the NLO perturbative calculation 
are the virtual one-loop contributions and the tree-graph (real emission) contributions. 
Since, at LO the relevant scalar products are $p_t\cdot s_t=0$, $p_t\cdot p_b=m_tE_b$ and 
$p_b\cdot s_t=\mp PE_b\cos\theta_P$ (in both helicity systems)
then the virtual one-loop corrections are contributed in the unpolarized rate ($\Gamma_A$) and 
the polar correlation function ($\Gamma_B$), which have been studied extensively before \cite{Nejad:2013fba}, 
while the azimuthal one ($\Gamma_C$) does not have any contribution from the virtual corrections.

The QCD NLO contribution results from the square of the real gluon emission graphs.
By working in the massless scheme where $m_b=0$, for the corresponding real amplitude squared one has
\begin{eqnarray}
|M^{\textbf{real}}|^2&=&-\frac{\pi^2C_F\alpha\alpha_s}{4\sin^2\theta_W}(-g^{\mu\nu}+\frac{p_W^\mu.p_W^\nu}{m_W^2})\times\nonumber\\
&&\bigg\{\frac{F_1}{(p_t\cdot p_g)^2}+\frac{F_2}{(p_b\cdot p_g)^2}-\frac{2F_3}{(p_t\cdot p_g)(p_b\cdot p_g)}\bigg\},\nonumber\\
\end{eqnarray}
where $C_F=4/3$ stands for the color factor, and
\begin{eqnarray}
F_1&=&Tr[\displaystyle{\not}p_b\gamma_\mu(1-\gamma_5)(m_t+\displaystyle{\not}p_t-\displaystyle{\not}p_g)
\gamma^\beta(\displaystyle{\not}p_t+m_t)\times\nonumber\\
&&(1+\gamma_5\displaystyle{\not}s_t)\gamma_\beta(m_t+\displaystyle{\not}p_t-\displaystyle{\not}p_g)(1+\gamma_5)\gamma_\nu],\nonumber\\
F_2&=&Tr[\displaystyle{\not}p_b\gamma^\beta(\displaystyle{\not}p_b+\displaystyle{\not}p_g)\gamma_\mu(1-\gamma_5)(m_t+\displaystyle{\not}p_t)
(1+\gamma_5\displaystyle{\not}s_t)\times\nonumber\\
&&(1+\gamma_5)\gamma_\nu(\displaystyle{\not}p_b+\displaystyle{\not}p_g)\gamma_\beta],\nonumber\\
F_3&=&Tr[\displaystyle{\not}p_b\gamma_\mu(1-\gamma_5)(m_t+\displaystyle{\not}p_t-\displaystyle{\not}p_g)\gamma^\beta
(m_t+\displaystyle{\not}p_t)\times\nonumber\\
&&(1+\gamma_5\displaystyle{\not}s_t)(1+\gamma_5)\gamma_\nu(\displaystyle{\not}p_b+\displaystyle{\not}p_g)\gamma_\beta].
\end{eqnarray}
In general, to regulate the gluon IR singularities we work in a D-dimensions approach, where
the differential decay rate for the real  contribution is given by
\begin{eqnarray}\label{aidaa}
d\Gamma=\frac{\mu_F^{2(4-D)}}{2m_t}|M^{\textbf{real}}|^2dR_3(p_t, p_b, p_g, p_{W}),
\end{eqnarray}
where, $\mu$ is an arbitrary reference mass and the 3-body phase space element $dR_3$ reads
\begin{eqnarray}
\frac{d^{D-1}\bold{p}_b}{2E_b}\frac{d^{D-1}\bold{p}_W}{2E_W}\frac{d^{D-1}\bold{p}_g}{2E_g}
(2\pi)^{3-2D}\delta^D(p_t-\sum_{g,b,W} p_f).\nonumber\\
\end{eqnarray}
Here, $d^{D-1}|\bold{p}|=|\vec{p}|^{D-2}d|\vec{p}|d\Omega$ where
the angular integral in D-dimensions will have to be written as
\begin{eqnarray}
\frac{d\Omega}{d\phi_P d\cos\theta_{P}}=-\frac{2\pi^{\frac{D-3}{2}}}{\Gamma(\frac{D-3}{2})}
(\sin\theta_{P})^{D-4}(\sin\phi_{P})^{D-4}.\nonumber\\
\end{eqnarray}
Considering the general form of the angular decay distribution (\ref{form}), 
as we showed in \cite{Nejad:2014sla} the unpolarized differential decay rate $d\Gamma_A/dx$ is independent of
the applied helicity system but the polar distribution of decay width $d\Gamma_B/dx$ 
depends on the various choices of possible coordinate systems, but all final results are free of IR singularities.
In the following we will concentrate on the differential azimuthal correlation function $d\Gamma_C/dx$, considering 
both helicity coordinate systems.\\
 In the system~1, the relevant scalar products are 
\begin{eqnarray}
\quad p_g\cdot s_t&=&-PE_g(\sin\theta_{gb}\sin\theta_P\cos\phi_P+\cos\theta_{gb}\cos\theta_{P}),\nonumber\\
p_g\cdot p_b&=&E_gE_b(1-\cos\theta_{gb}),\nonumber\\
p_b\cdot s_t&=&-PE_b\cos\theta_{P},
\end{eqnarray}
where $\theta_{gb}$ is the polar angle between the gluon and the bottom quark momenta in the event plane, so
$\cos\theta_{gb}=(m_t^2-m_W^2-2m_t(E_b+E_g)+2E_b E_g)/(2E_b E_g)$.
To calculate the $d\Gamma_C/dx_b$, in (\ref{aidaa}) we fix the momentum of the b-quark and integrate over the
energy of the gluon, which ranges as $m_t S (1-x_b)\leq E_g\leq m_t S(1-x_b)/(1-2Sx_b)$, where $S=(1-\omega)/2$.
Therefore, one has
\begin{eqnarray}\label{fir}
\frac{d\Gamma_{1C}}{dx_b}&=&\Gamma_B^{\textbf{(0)}}\frac{\alpha_s C_F}{2(1-\omega)(1-2\omega)}
\bigg\{4(\omega-1)x_b-2\omega^2-5\omega\nonumber\\
&&+11-\frac{4(1+\omega)}{x_b}+\frac{1}{(1-x_b(1-\omega))^{\frac{3}{2}}}\Big[-16+\nonumber\\
&&2\omega+10\omega^2+\frac{4(1+\omega)}{x_b}+2(1-\omega)^2(\omega-5)x_b^2+\nonumber\\
&&(1-\omega)^3x_b^3+(7\omega^3-3\omega^2-25\omega+21)x_b\Big]\bigg\},\nonumber\\
\end{eqnarray}
where $x_b$ is defined in (\ref{variable}).
This result can be compared against known results presented in \cite{Fischer:2001gp} after integrating over 
$x_b (0\leq x_b\leq 1)$.\\
Since the observed mesons in top quark decays can be also produced through  a fragmenting real gluon, therefore, to obtain 
the most accurate energy distribution of the B meson one has to add the contribution of gluon fragmentation to 
the b-quark one. In \cite{Nejad:2013fba}, it is shown that the gluon contribution can be 
important at a low energy of the detected meson so that this contribution decreases the size of decay rate at the threshold energy. 
$d\Gamma_A/dx_g$ is the same in both coordinate systems and can be found in \cite{Kniehl:2012mn}, 
and the analytical expression for the $d\Gamma_{1B}/dx_g$ in the helicity system~1 
is presented in  \cite{Nejad:2013fba} and the $d\Gamma_{2B}/dx_g$ is given in \cite{Nejad:2014sla}, 
where $x_g$ is defined in (\ref{variable}). In the coordinate system~1, the azimuthal differential width $d\Gamma/dx_g$ reads
\begin{eqnarray}
\frac{d\Gamma_{1C}}{dx_g}&=&\Gamma_B^{\textbf{(0)}}\frac{\alpha_s C_F}{(1-\omega)(1-2\omega)}
\bigg\{4(1-\omega)x_g-8+6\omega+\nonumber\\
&&\frac{2(1+\omega)}{x_g}+\frac{1-2\omega}{x_g^2}+\frac{1}{4(1-x_g(1-\omega))^{\frac{3}{2}}}\Big[\nonumber\\
&&-7(1-\omega)^3x_g^3-2(1-\omega)^2(5\omega-19)x_g^2-\nonumber\\
&&(9\omega^3+19\omega^2-93\omega+65)x_g-\frac{4(1-2\omega)}{x_g^2}+\nonumber\\
&&2(2\omega^3-13\omega^2-5\omega+20)+\frac{2(6\omega^2-13\omega-1)}{x_g}\Big]\bigg\}.\nonumber\\
\end{eqnarray}
In the helicity coordinate system~2, the relevant scalar products are 
\begin{eqnarray}
\quad p_g\cdot s_t&=&PE_g(\sin\theta_{gW}\sin\theta_P\cos\phi_P-\cos\theta_{gW}\cos\theta_{P}),\nonumber\\
\quad p_b\cdot s_t&=&PE_b(\sin\theta_{bW}\sin\theta_P\cos\phi_P-\cos\theta_{bW}\cos\theta_{P}),\nonumber\\
p_g\cdot p_W&=&m_t(E_W+E_g)-\frac{m_t^2+m_W^2}{2},\nonumber\\
p_b\cdot p_W&=&m_t(E_W+E_b)-\frac{m_t^2+m_W^2}{2},\nonumber\\
p_W\cdot s_t&=&-P|\vec{p}_W|\cos\theta_{P},
\end{eqnarray}
and $p_t\cdot s_t=0$. In the system~2, $\theta_{bW}$ is the polar angle between the b-quark momentum and the W boson ($z$-axis)
and $\theta_{gW}$ is the angle between the gluon and the W boson, whereas $\cos\theta_{gW}=(m_t^2+m_W^2-2m_t(E_W+E_g)+2E_W E_g)/(2E_g p_W)$
and $\cos\theta_{bW}=(m_t^2+m_W^2-2m_t(E_b+E_W)+2E_b E_W)/(2E_b p_W)$ with $p_W=\sqrt{E_W^2-m_W^2}$.\\
As before, to calculate the azimuthal differential rate $d\Gamma_C/dx_b$, in (\ref{aidaa})
we fix the momentum of the b-quark but we integrate over the energy of the
$W$ boson, which ranges as $m_t(\omega+[1-2Sx_b]^2)/(2(1-2Sx_b))\leq E_g\leq m_t (1-S)$.
Therefore, in the coordinate system~2 the azimuthal differential width $d\Gamma_C/dx_b$ is expressed as
\begin{eqnarray}\label{sec}
\frac{d\Gamma_{2C}}{dx_b}&=&\Gamma_B^{\textbf{(0)}}\frac{\alpha_s C_F}{2(2-(1-\omega)x_b)^2(1-\omega)(1-2\omega)}\times\nonumber\\
&&\bigg\{x_b^2(1-\omega)^2(2\omega^2-3\omega-3)+4\omega^3+22\omega^2-16\omega\nonumber\\
&&-2+2x_b(1-\omega)(-6\omega^2+7\omega+3)-\nonumber\\
&&2\sqrt{\frac{\omega+(1-x_b(1-\omega))^2}{1+\omega}}\Big[x_b^2(1-\omega)^2(2\omega-1)+\nonumber\\
&&\hspace{-0.5cm}2x_b(\omega-1)(2\omega^2+4\omega-1)+(1+\omega)(2\omega^2+9\omega-1)\Big]\nonumber\\
&&+8\omega(2-x_b(1-\omega))^2\sqrt{1-x_b(1-\omega)}\bigg\},
\end{eqnarray}
and for the gluon one, we have
\begin{eqnarray}\label{sec2}
\frac{d\Gamma_{2C}}{dx_g}&=&\Gamma_B^{\textbf{(0)}}\frac{\alpha_s C_F}{(1-\omega)(1-2\omega)}\bigg\{
\frac{1}{2x_g\sqrt{1-x_g(1-\omega)}}\Big[\nonumber\\
&&\omega(1-x_g^2)-(1-2\omega^2)(1-x_g)^2\Big]+\nonumber\\
&&\Big(\sqrt{\frac{\omega+(1-x_g(1-\omega))^2}{1+\omega}}-\sqrt{1-x_g(1-\omega)}\Big)\nonumber\\
&&\times\Big[2\omega-1-\frac{(1+\omega)^2(1-2\omega)}{x_g^2(1-\omega)^2}+\frac{2(1+2\omega^2)}{x_g(1-\omega)}\Big]\bigg\}.\nonumber\\
\end{eqnarray}

\section{Numerical analysis}
\label{sec:three}

By determining the differential decay rates in the parton level, in  the first step
we turn to our numerical predictions  
of the unpolarized and polarized decay rates by integrating 
$d\Gamma/dx_b$ over $x_b (0\leq x_b\leq 1)$, 
while the strong coupling constant is evolved from  $\alpha_s(m_Z)=0.1184$ to $\alpha_s(m_t)=0.1070$.
By combining our results for the differential azimuthal correlation 
functions (Eqs.~(\ref{fir}) and (\ref{sec})) with the results obtained for the
unpolarized rate \cite{Kniehl:2012mn} and the polar correlation rate in the system~1  \cite{Nejad:2013fba},
one has
\begin{eqnarray}
\frac{d\Gamma_1^{\textbf{NLO}}}{d\phi_{1P} d\cos\theta_{1P}}&=&\frac{1}{4\pi}
\bigg\{\Gamma_A^{\textbf{(0)}}(1-0.08542)\nonumber\\
&&-\Gamma_B^{\textbf{(0)}}(1-0.1303)P\cos\theta_{1P}\nonumber\\
&&-\Gamma_B^{\textbf{(0)}}(0.0980)P\sin\theta_{1P}\cos\phi_{1P}\bigg\}\nonumber\\
&&=\frac{\Gamma_A^{\textbf{NLO}}}{4\pi}
\Big\{1-0.3777 P\cos\theta_{1P}\nonumber\\
&&-0.0426P\sin\theta_{1P}\cos\phi_{1P}\Big\},
\end{eqnarray}
and by considering the polar correlation one in the system~2 \cite{Nejad:2014sla}, one has
\begin{eqnarray}
\frac{d\Gamma_2^{\textbf{NLO}}}{d\phi_{2P} d\cos\theta_{2P}}&=&\frac{1}{4\pi}
\bigg\{\Gamma_A^{\textbf{(0)}}(1-0.08542)\nonumber\\
&&+\Gamma_B^{\textbf{(0)}}(1-0.2814)P\cos\theta_{2P}\nonumber\\
&&+\Gamma_B^{\textbf{(0)}}(0.01446)P\sin\theta_{2P}\cos\phi_{2P}\bigg\}\nonumber\\
&&=\frac{\Gamma_A^{\textbf{NLO}}}{4\pi}
\Big\{1+0.3121\cos\theta_{1P}\nonumber\\
&&+0.0063P\sin\theta_{1P}\cos\phi_{1P}\Big\},
\end{eqnarray}
where $\Gamma_A^{\textbf{(0)}}=1.4705$ and $\Gamma_B^{\textbf{(0)}}=0.5841$ if one sets $m_W=80.399$~GeV,
$m_t=172.98$~GeV and  $\sin^2\theta_W=0.2312$ \cite{Nakamura:2010zzi}.\\
As it is seen, the azimuthal correlation generated by the radiative corrections is 
quite small, especially in the second coordinate system. We can assert that,
if top quark decays reveal a violation of the SM (V-A) current structure in the
azimuthal correlation function which exceeds the $5\%$ level at the system~1 and the $1\%$ level at the system~2,
the violation must have a non-SM origin.
\begin{figure}
\begin{center}
\includegraphics[width=0.9\linewidth,bb=37 192 552 629]{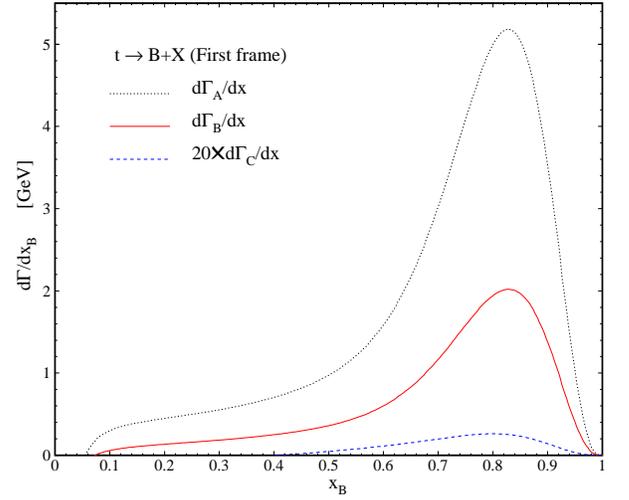}
\caption{\label{fig1}%
$x_B$ distribution of $d\Gamma^{\textbf{NLO}}/dx_B$ in the helicity system~1, considering the unpolarized (dotted line),
the polar (solid line) and the azimuthal (dashed line) contributions.}
\end{center}
\end{figure}
\begin{figure}
\begin{center}
\includegraphics[width=0.9\linewidth,bb=37 192 552 629]{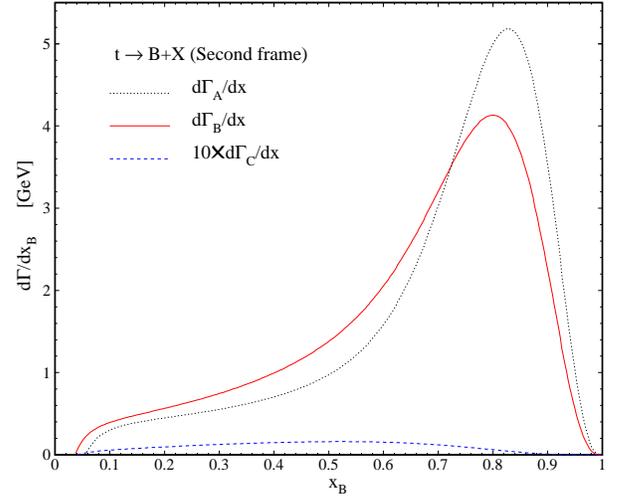}
\caption{\label{fig2}%
As in Fig~\ref{fig1}, but  in the helicity system~2.}
\end{center}
\end{figure}

In the last step, we present our phenomenological results for the energy spectrum $d\Gamma/dx_B$ of the
B meson, where we define the normalized energy fraction of the B meson 
as $x_B=2E_B/(m_t(1-\omega))$ (\ref{variable}). According to the the factorization theorem of the 
QCD-improved parton model \cite{jc}, the energy distribution of B meson
can be expressed as the convolution of the parton-level spectrum $d\Gamma/dx_i$  with the
nonperturbative fragmentation functions  $D_i^B(z,\mu_F)$ as
\begin{eqnarray}\label{convolute}
\frac{d\Gamma}{dx_B}=\sum_{i=b, g}\frac{d\Gamma}{dx_i}(\mu_R, \mu_F)\otimes D_i^B(\frac{x_B}{x_i},\mu_F).
\end{eqnarray}
The integral convolution is defined as $(f\otimes g)(x)=\int_x^1 dx f(z) g(x/z)$.
In (\ref{convolute}), $\mu_F$ and $\mu_R$ are the factorization and the renormalization scales, respectively,
that the scale $\mu_R$ is associated with the renormalization of the strong coupling constant and
a choice often made consists of setting $\mu_R=\mu_F$. As in our previous works,
we adopt the convention $\mu_R=\mu_F=m_t$.\\
In (\ref{convolute}), $D_{i}(z, \mu_F)$ is the nonperturbative FF describing the 
transition $(b, g)\rightarrow B$ which is process independent. 
Several models are proposed to describe the nonperturbative transition from
a parton into a hadron state. Here, following Ref.~\cite{Kniehl:2008zza}
we employ the B meson FF determined at NLO in the ZM-VFN scheme and 
obtained by fitting the experimental data from the 
ALEPH  and OPAL collaborations at CERN LEP1 and by SLD at SLAC SLC.
Authors in \cite{Kniehl:2008zza} have parametrized the $z$ distribution of the $b\rightarrow B$ FF
at the initial scale $\mu_0=m_b$ as  $D_b^B(z, \mu_0)=Nz^\alpha(1-z)^\beta$ (power model), 
while the gluon FF is set to zero at the starting scale and is evolved to higher scales using the DGLAP equations \cite{dglap}.
Their results for the fit parameters at the initial scale are $N=4684.1,\alpha=16.87$ and $\beta=2.628$
with $\overline{\chi^2}=1.495$.  
Following Ref.~\cite{Nakamura:2010zzi}, as numerical input values we take
$m_W=80.339$ GeV,
$m_b=4.78$~GeV, 
$m_B=5.279$~GeV
and the typical QCD scale $\Lambda_{\overline{MS}}^{(5)}=231$~MeV.
Note that, in the ZM-VFN scheme the b-quark mass only enter through the initial condition of the  FF.\\
To study the $x_B$ scaled energy  distributions of B mesons, we consider the quantity
$d\Gamma(t(\uparrow)\to B+X)/dx_B$ in the two helicity coordinate systems.
In \cite{Kniehl:2012mn,Nejad:2013fba}, we showed that the $g\rightarrow B$ contribution into the NLO energy spectrum of the 
B-meson is negative and appreciable only
in the low-$x_B$ region and for higher values of $x_B$ the NLO result is
practically exhausted by the $b\rightarrow B$ contribution. The contribution of the gluon is calculated to see where
it contributes to $d\Gamma/dx_B$ and can not be discriminated in the meson spectrum  as an experimental quantity.
In the scaled energy of mesons, all contributions
including the bottom quark, gluon and light quarks contribute.
In Fig.~\ref{fig1}, the $x_B$-spectrum of the B meson produced through the unpolarized  top quark decay (dotted line) is shown.
The polar (solid line) and the azimuthal (dashed line) contributions in the helicity system~1  are also studied. 
As we explained, the azimuthal correlation is prohibited at LO, which explains the smallness of the corresponding result.
Note that  the threshold occurs at $x_B\geq 2m_B/(m_t(1-\omega))=0.078$.
In Fig.~\ref{fig2}, the same predictions are shown in the helicity system~2.

\section{Conclusions}
\label{sec:four}
 
To study the ${\cal O}(\alpha_s)$ spin-dependent energy spectrum of hadrons produced from polarized top quark decays, 
one needs to know the NLO radiative corrections to the angular differential decay rates of the process
$t(\uparrow)\rightarrow b+W^+$. In our previous works, the unpolarized decay rate ($d\Gamma_A/dx_i$) and the polar correlation 
one ($d\Gamma_B/dx_i$) were calculated at the parton-level in two different helicity coordinate systems.
These various helicity systems provide independent probes of the polarized top quark decay dynamics.\\
Here, by considering two helicity systems
we have calculated the ${\cal O}(\alpha_s)$ corrections to the differential azimuthal correlation function ($d\Gamma_c/dx_i$)
which vanishes at the Born term level. These quantities are required to calculate
the $x_B$ distribution $d\Gamma_C/dx_B$ of $t\rightarrow B+X$.
Comparing future measurements of the polarized and unpolarized partial widths $d\Gamma/dx_B$
at the LHC with our NLO predictions, one will be able to test the universality and scaling
violations of the B meson FFs. These measurements will finally be the primary source of information
on the B meson FFs and the azimuthal correlation function $d\Gamma_C/dx_B$ can also constrain the $g\rightarrow B$
and $b\rightarrow B$ FFs even further.\\
We also found that the ${\cal O}(\alpha_s)$ radiative corrections
to the azimuthal correlation observable $\Gamma_C$ are so small, especially in the helicity system~2.
Specifically, it is safe to say that,
if top quark decays reveal a violation of the SM (V-A) current structure in the
azimuthal correlation function which exceeds the $5\%$ level at the system~1,
the violation must have a non-SM origin.

\end{document}